\documentclass[letter]{aa} 

\usepackage{graphicx}
\usepackage{txfonts}
\usepackage{orcidlink}

\begin{document} 

\title{The uncertain interstellar medium of high-redshift quiescent galaxies: 
Impact of methodology}

\author{R. Gobat\orcidlink{0000-0003-0121-6113}\inst{1}
\and C. D'Eugenio\orcidlink{0000-0001-7344-3126}\inst{2,3}
\and D. Liu\orcidlink{0000-0001-9773-7479}\inst{4}
\and G.B. Caminha\orcidlink{0000-0001-6052-3274}\inst{5,6}
\and E. Daddi\orcidlink{0000-0002-3331-9590}\inst{7}
\and D. Bl\'{a}nquez\inst{8}
}

\institute{
Instituto de F\'{i}sica, Pontificia Universidad Cat\'{o}lica de Valpara\'{i}so, 
Casilla 4059, Valpara\'{i}so, Chile
\and Instituto de Astrof\'{i}sica de Canarias (IAC), E-38205 La Laguna, Tenerife, 
Spain
\and Universidad de La Laguna, Dpto. Astrof\'{i}sica, E-38206 La Laguna, Tenerife, Spain
\and Max-Planck-Institut f\"{u}r extraterrestrische Physik, Gie{\ss}enbachstra{\ss}e 1, 
D-85748 Garching, Germany
\and Technische Universit\"{a}t M\"{u}nchen, Physik-Department, James-Franck Str.
1, 85748 Garching, Germany
\and Max-Planck-Institut f\"{u}r Astrophysik, Karl-Schwarzschild-Str. 1, D-85748 Garching, 
Germany
\and CEA, Irfu, DAp, AIM, Universit\'{e} Paris-Saclay, Universit\'{e} de Paris, CNRS, 
F-91191 Gif-sur-Yvette, France
\and DTU-Space, Technical University of Denmark, Elektrovej 327, DK-2800 Kgs. Lyngby, 
Denmark
}

\date{}

 
\abstract{
How much gas and dust is contained in high-redshift quiescent galaxies (QGs) is currently 
an open question with relatively few and contradictory answers, as well as important 
implications for our understanding of the nature of star formation quenching processes at 
cosmic noon. Here we revisit far-infrared (FIR) observations of the REQUIEM-ALMA sample 
of six $z=1.6-3.2$ QGs strongly lensed by intermediate-redshift galaxy clusters. 
We measured their continuum emission using priors obtained from high resolution near-infrared 
(NIR) imaging, as opposed to focusing on point-source extraction, converted it into dust masses 
using a FIR dust emission model derived from statistical samples of QGs, and compared the 
results to those of the reference work.
We find that, while at least the most massive sample galaxy is indeed dust-poor, the picture 
is much more nuanced than previously reported. In particular, these more conservative 
constraints remain consistent with high dust fractions in early QGs. We find that these 
measurements are very sensitive to the adopted extraction method and conversion 
factors: the use of an extended light model to fit the FIR emission increases 
the flux of detections by up to 50\% and the upper limit by up to a factor 6. Adding the 
FIR-to-dust conversion, this amounts to an order of magnitude difference in dust fraction, 
casting doubts on the power of these data to discriminate between star 
formation quenching scenarios. Unless these are identified by other means, mapping the dust 
and gas in high-redshift QGs will continue to require somewhat costly observations. 
}

\keywords{Galaxies:early-type -- Galaxies:formation -- Galaxies:ISM -- Gravitational 
lensing:strong}

\titlerunning{Re-REQUIEM}
\authorrunning{Gobat et al.}

\maketitle
%

\section{\label{intro}Introduction}

In the local Universe, most of the stars are held in so-called quiescent galaxies, which 
are predominantly spheroidal systems where star formation activity is at very low levels 
or entirely absent, and which have evolved passively, only becoming older and redder for 
the past ten billion years. Unsurprisingly, contemporary quiescent galaxies by and large 
lack the cold gas that fuels star-forming (SF) galaxies such as the Milky Way 
\citep[e.g.,][]{Sai11,You11,Bos14}. Whether they always did, on the other hand, 
remains an unsolved yet deceptively important question. 
The stellar component of quiescent galaxies (QGs) has been extensively studied and its 
evolution traced out to $z\sim4$, both photometrically \citep[e.g.,][]{Dav17} and 
spectroscopically \citep{Gla17,Sch18,Tan19,Val20}. However, the various mechanisms that 
have been put forward to quench star formation, by either preventing the cooling of gas 
in and onto galaxies \citep[e.g.,][]{BD03,Cro06,Cat06}, stabilizing it \citep{Mrt09}, 
or outright expelling it \citep[e.g.,][]{DiM05,Hop06}, might not leave sufficiently 
conspicuous signatures in stellar populations. Indeed, galactic archaeology by way of 
spectroscopic modeling \citep[e.g.,][]{Ono15,Gob17,Val20} has so far yielded only 
circumstantial evidence on the quenching pathways of these galaxies 
\citep[e.g.,][]{Ono15,MB18,Paw19,Bel19}, constraining the timescale of quenching, 
but not its specific mechanism.
On the other hand, the state of the interstellar medium (ISM) of QGs after quenching 
should  be sensitive to the specific scenario that led to quiescence: for example, a 
completely expulsive quenching naturally leaves less gas than gravitational stabilization 
or a mechanism that heats the ISM which, in the latter case, is in a warmer phase (e.g., 
neutral) on average than if stabilized against fragmentation.\\

While the gas content of normal, SF galaxies has been extensively traced up to 
$z\sim4$~\citep[e.g.,][]{Tac18,Tac20,Liu19}, similar efforts targeting QGs remain 
sparser and heterogeneous, with CO observations generally placing constraints on 
their molecular gas fraction ($f_{\text{H}_2}$) ranging from $f_{\text{H}_2}\lesssim$6\% 
upper limits \citep{Sar15,Bez19,Wil21} to $f_{\text{H}_2}=15-25$\% in some outliers 
\citep{Rud17,Hay18}. 
Interstellar dust, which correlates with total gas \citep[e.g.,][]{Mag12}, has so far 
provided the only statistical, though more indirect, constraints on the ISM of QGs for 
the first five billion years of the history of the Universe. Two approaches based on 
dust emission have been used which, however, appear to yield contradictory results: 
the first, using far-infrared (FIR) spectral energy distributions (SEDs) averaged over 
large samples, suggests that $z>1$~QGs retain relatively large amounts of dust and gas 
\citep[hereafter, G18 and M21]{Gob18,Mag21}, corresponding to gas fractions 
of $f_{\text{gas}}\sim7$\% at solar metallicity. The data on which these SEDs are 
based have a variable resolution that is typically lower than the angular size of 
individual galaxies. They might therefore include contributions from nearby sources, 
which have to be corrected for. The second approach, on the other hand, involves 
resolved observations of strongly lensed (and thus magnified) QGs with, however, 
more limited statistics. Based on these, \citet[hereafter, W21]{Whi21a} derived 
constraints on dust masses which imply significantly lower upper limits of 
$0.1-1$\% on $f_{\text{gas}}$. However, these conclusions are also based on 
assumptions, such as the compactness of emission, conversion factors, and (in 
particular, since the observations only sample the Rayleigh-Jeans tail of 
dust emission) dust temperature. It is therefore not immediately clear whether 
the apparent tension between these two types of constraints is not simply an 
artifact of mismatched assumptions.\\

To remedy this doubt, we revisited the lensed QG sample of W21 using a methodology 
consistent with that of G18 and M21. Section~\ref{data} briefly summarizes the 
sample and data, Sect.~\ref{fit} describes our analysis, Sect.~\ref{results} provides 
the results, and Sect.~\ref{conclusion} presents our conclusions. 
We assumed a concordance cosmology with $H_0=70$\,km\,s$^{-1}$\,Mpc$^{-1}$, 
$\Omega_{\text{M}}=0.3$, and $\Omega_{\Lambda}=0.7$, as well as a \citet{Sal55} 
initial mass function (IMF), with quantities from the literature having been 
converted to this IMF.

\section{\label{data}Sample and data}

The REQUIEM-ALMA sample (W21) consists of six QGs at $z=1.5-3.2$ observed under 
programs 2018.1.00276 (PI K.E. Whitaker) and 2019.1.00227 (PI J. Tan) with 
the Atacama Large Millimeter/Submillimeter Array (ALMA) at 1.3\,mm, corresponding 
to the ALMA Band 6 and $300-500$\,$\mu$m rest frame. 
We obtained the raw data from the ALMA Science Archive and processed them to produce 
calibrated visibility sets, which were then split and binned in both time and channels, 
and finally exported as a single continuum UVFITS table per QG with 
\textsc{GILDAS}\footnote{\texttt{http://www.iram.fr/IRAMFR/GILDAS}}. 
All six QGs are strongly lensed by intermediate-redshift ($z=0.3-0.8$) massive galaxy 
clusters, with two creating giant arc images. Due to the nature of the lenses, they 
benefit from optical-near-infrared (NIR) imaging by the Hubble (HST) and \emph{Spitzer} space 
telescopes, accumulated over the years under a variety of observing programs and 
publicly available in their respective archives. Table~\ref{tab:hst} summarizes 
the photometric coverage of these fields. The lensed galaxies themselves have also been 
observed spectroscopically, with them and the data being described in 
\citet[hereafter, N18]{New18} and \citet{Man21}. We refer the reader to these 
articles for more information and we adopt their naming conventions throughout. 

\begin{figure*}
\centering
\includegraphics[width=1\textwidth]{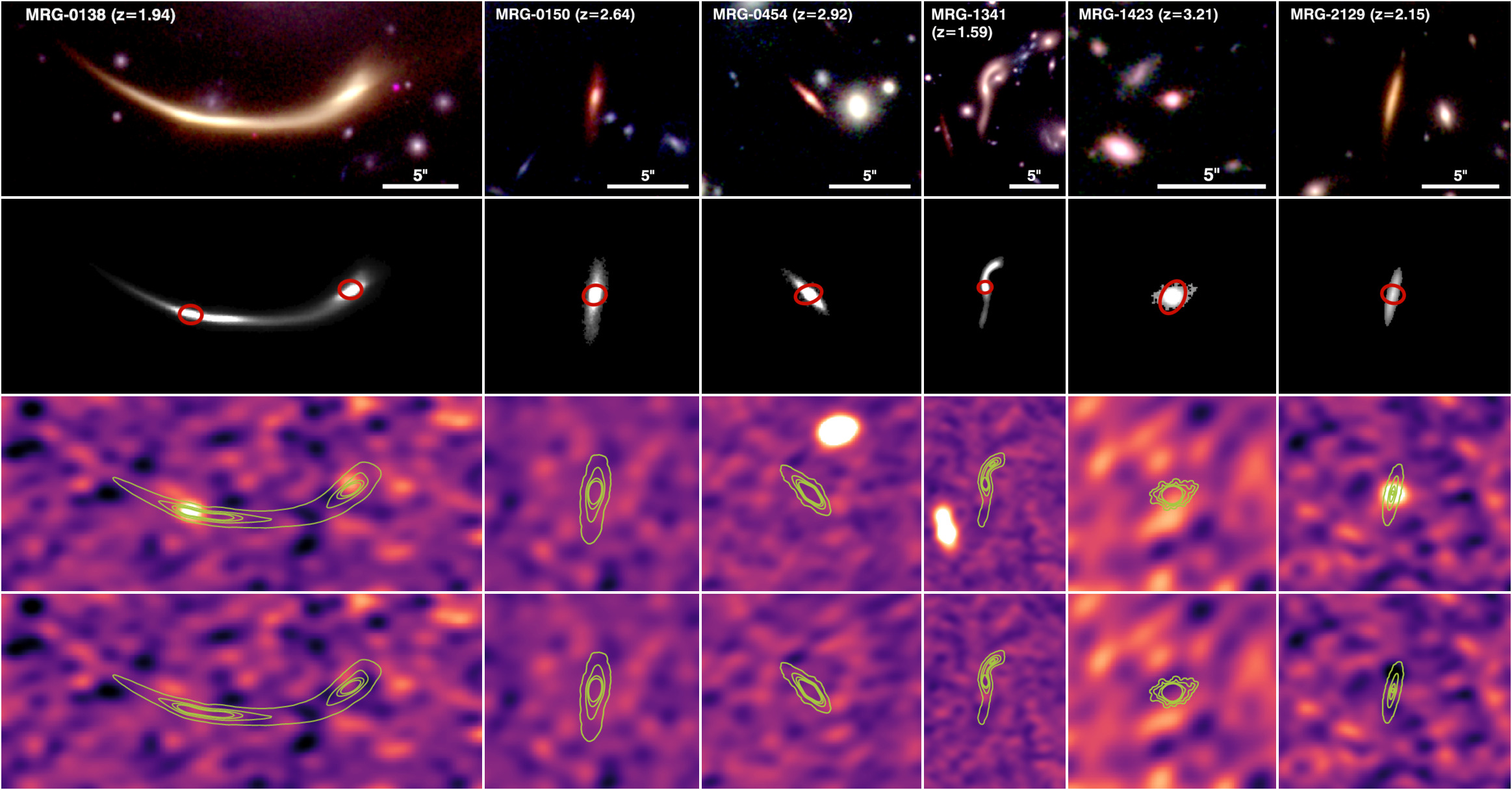}
\caption{Cutouts of the six lensed QGs reanalyzed in this work. The rows, from top to 
bottom, show: color images (HST F160W, F140W, or F125W, and F105W) with scale 
bar; stellar light models with the ALMA synthesized beam shown as a red ellipse; and 
ALMA Band 6 continuum images and residuals after subtraction of the model, with the 
1.6\,$\mu$m light distribution shown as green contours.
}
\label{fig:cutouts}
\end{figure*}

\section{\label{fit}Analysis}

Based on two point-like detections, W21 concluded that the 1.3\,mm emission from dust was 
confined to compact, unresolved regions of the QGs. These correspond, spatially, to their 
cores (the central kiloparsec, in the case of the most extended, MRG-0138), which are 
also the brightest regions of the galaxies in stellar light. More recently, the QG 
MRG-2129 was followed up on with ALMA at very high resolution by \citet{Mor22}, who 
similarly concluded that the emission is compact and localized. However, this new 
observation consists of a single pointing with a maximum recoverable scale of 
$\sim$0.6'', or half the beam width of the data used in W21. 
Therefore, if conversely the distribution of dust includes an overall continuous 
component following the stellar one, the emission from lower-surface brightness 
regions could very well be either below the noise level of the data or missed 
entirely by the instrument. 
As a test, we examined resolved, integral field spectroscopy of two of the most 
extended galaxies in the sample, MRG-0138 and MRG-1341 (see Appendix~\ref{appendix:muse}). 
Within the wavelength range of the spectra, we found no appreciable difference between 
the bright central regions of these galaxies and their extended outer regions, which 
would for example suggest that one contains a larger fraction of younger stars than 
the former.\\

For each galaxy, we first constructed a model of their stellar distribution from the 
longest-wavelength high-resolution data available, namely the HST Wide Field 
Camera 3 (WFC3) F160W imaging, corresponding to the rest-frame optical range. Cutouts 
centered on the QGs were extracted from the F160W images and the light of nearby 
objects was modeled and subtracted using \textsc{galfit} \citep{Peng10}. All pixels below 
3$\sigma$~of the background value were then zeroed out, using \textsc{SExtractor} 
\citep{BA96} on the residual image to isolate the QG. We note that, while this 
threshold was chosen to avoid including image noise in the model, it might cut out 
some of the galaxies' extended profiles. However, as the F160W data are deep compared 
to the brightness of the lensed QGs, the loss of emission should not be significant. 
The use of these stellar light models should therefore come closest to reproducing 
the effective aperture of the SED stacks in M21, minus the potential residual 
contamination by nearby sources. The cutout images and resulting models are shown 
in Fig.~\ref{fig:cutouts} (top and second row, respectively).\\

To account for astrometry differences between the F160W and ALMA data, the fit allowed 
for a free offset. For detected objects, this yielded offsets of $<$0.15", which 
is significantly smaller than the spatial resolution of the ALMA data. 
Any bright sources present in the field of view (MRG-0454 and MRG-1341) were modeled 
at the same time, so as to mitigate the possibility of contamination from sidelobes. 
In cases where the residuals of the extended model fit contained noticeable 
($>$3$\sigma$) leftover emission, we also allowed for excess central emission in the 
form of an additional point source with free amplitude for a total of 
two components per fit in this case. 
One exception was MRG-0138, where the lensed arc targeted by the ALMA observation 
is a combination of two images bisected by a critical line. We did not attempt to 
separate them, but rather counted them as a single image, thus requiring the use of 
two point-source components, with relative amplitudes fixed by the ratio of mean 
magnifications between the two images given in N18. 
We then corrected the quantities derived for this object to the magnification factor 
of the first image (N18). As a sanity check and to allow for a direct comparison with 
W21, we also extracted 1.3\,mm fluxes by fitting a point source without the total 
light model. Uncertainties (and thus upper limits) for model and point source fluxes were estimated 
from the r.m.s. dispersion of a thousand extractions performed with each component at 
random positions with large offsets from phase center (that is, staying well clear of 
the target galaxy and other sources within the field of view). The resulting magnified 
fluxes and errors, or 3$\sigma$~upper limits, are shown in Table~\ref{tab:results}.\\

\subsection{\label{mdust}Dust masses}

We then converted these fluxes into (magnified) dust masses using the average FIR 
SED template for $z>0$~QGs of M21, sampled at 1.3\,mm. This model has a dust 
temperature of $T_d=21$\,K, which is a few Kelvin lower than the mass-weighted 
$T_d$~typically used for SF galaxies \citep[hereafter S16]{Sco16,Sco17}. In addition 
to being consistent with the methodology of M21, with which we compare in this study, 
we consider it a more likely description of the ISM of QGs which, given their 
quiescence, should be subjected to a softer radiation field than in SF galaxies. 
If the FIR SEDs used in M21 were contaminated by, for example, SF occurring at a similar 
redshift (i.e., from satellite galaxies), the true $T_d$~of high-redshift QGs might 
be even lower. To summarize, here, we consider three analysis methods. In order of 
complexity, they are as follows: point-source extraction converted to a dust mass using S16, 
point-source extraction converted to dust mass using the M21 SED, and an extended 
model fit converted to dust using M21. For clarity, we do not include the 
fourth possible combination here, that is, the extended model fit with the S16 calibration, 
as it falls between the first and third method and can be easily inferred from 
the first three (see Sect.~\ref{results}). The dust masses or upper limits 
corresponding to both model+point and point-source fluxes are given in 
Table~\ref{tab:results}.\\

\subsection{\label{mstar}Stellar masses}

For consistency with the ALMA analysis, we also recomputed stellar masses for 
the six REQUIEM-ALMA QGs from SEDs based on the F160W light model used to fit 
the ALMA data. That is, we extracted photometry from HST images using 
\textsc{SExtractor} in dual-image mode, with the F160W model as a reference. For 
\emph{Spitzer} data, which have significantly lower resolution and are thus 
often blended, we instead performed a multicomponent 
decomposition\footnote{with \textsc{MPFIT} \citep{MPF}} of the image, using the 
F160W model and \textsc{galfit} models of nearby objects convolved with the point 
response functions of \emph{Spitzer}/IRAC. We then modeled these SEDs with 
\citet{BC03} stellar population templates, assuming star formation histories 
(SFHs) with exponential cutoffs, as already used for high-redshift QGs 
\citep[e.g.,][]{Sch18,Val20}. We checked that the stellar masses thusly obtained 
are consistent with those from the literature. We refer the reader to 
Appendix~\ref{appendix:masses} for more details.

\begin{table*}
\caption{Lensed galaxy properties}
\renewcommand{\arraystretch}{1.5}
\setlength{\tabcolsep}{10pt}
\begin{tabular}{c c c c c c c c}
\hline\hline
ID & $z$ & log\,$\mu$M$_{\star}$ & $\mu f_{\text{1.3mm,model}}$ & 
$\mu f_{\text{1.3mm,point}}$ & log\,$\mu$M$_{d,\text{model}}$ & 
log\,$\mu$M$_{d,\text{point}}$ & $\mu$\\
& & M$_{\odot}$ & $\mu$Jy & $\mu$Jy & M$_{\odot}$ & M$_{\odot}$ & \\
\hline
0138 & 1.9439$^{\text{a}}$ & 13.05$^{+0.14}_{-0.10}$ & $352\pm116$ & $285\pm29$ & 
$9.11\pm0.15$ & $9.02\pm0.04$ & $12.5\pm5.4^{\text{a}}$\\
0150 & 2.6355$^{\text{a}}$ & 12.24$^{+0.25}_{-0.19}$ & $<$180 & $<52$ & $<9.30$ & 
$<8.76$ & $4.4\pm1.1^{\text{a}}$\\
0454 & 2.9225$^{\text{b}}$ & 11.72$^{+0.13}_{-0.09}$ & $<174$ & $<42$ & $<8.81$ & 
$<8.19$ & $10.9\pm2.1^{\text{b}}$\\
1341 & 1.5943$^{\text{b}}$ & 12.20$^{+0.14}_{-0.12}$ & $<161$ & $<27$ & $<8.76$ & 
$<7.98$ & $30\pm8^{\text{b}}$\\
1423 & 3.2092$^{\text{b}}$ & 11.27$^{+0.21}_{-0.07}$ & $<91$ & $28\pm11$ & 
$<8.52$ & $8.01\pm0.17$ & $2.7\pm2^{\text{b}}$\\
2129 & 2.1487$^{\text{a}}$ & 11.88$^{+0.14}_{-0.08}$ & $247\pm66$ & $164\pm17$ & 
$8.96\pm0.12$ & $8.78\pm0.05$ & $4.6\pm0.2^{\text{b}}$\\
\hline
\end{tabular}
$^{\text{a}}$\citet{New18}, $^{\text{b}}$\citet{Man21}
\label{tab:results}
\end{table*}

\section{\label{results}Results}

Of the two point-like detections in W21, the emission of MRG-2129 is well-fitted 
by the extended stellar light model (Fig.~\ref{fig:cutouts}, right) without 
compact emission, with slightly lower residuals than when using a point source; 
although, the difference is not significant. With the same conversion factor, 
this yields a 50\% larger $M_d$~than when only considering compact emission 
\citep[W21,][]{Mor22}. 
On the other hand, MRG-0138 requires an additional point-like component, 
accounting for $\sim$45\% of the total flux per image. For this object, the 
error on the flux is thus taken as the quadrature sum of model and point-source 
uncertainties. The other four galaxies in the sample remain undetected at 
the 3$\sigma$~level, in which case we only considered uncertainties from 
the stellar light model.\\

\begin{figure*}
\centering
\includegraphics[width=0.495\textwidth]{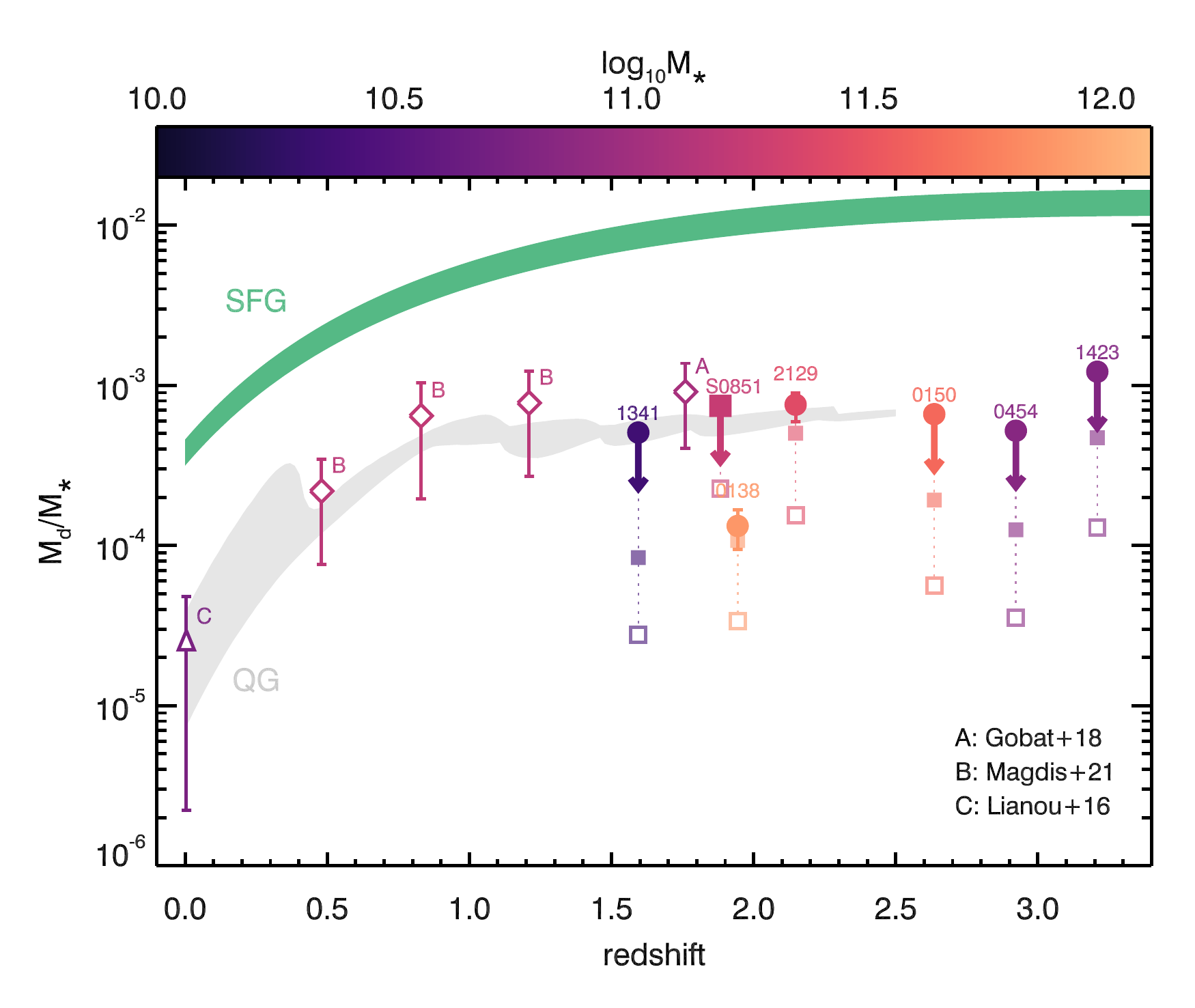}
\includegraphics[width=0.495\textwidth]{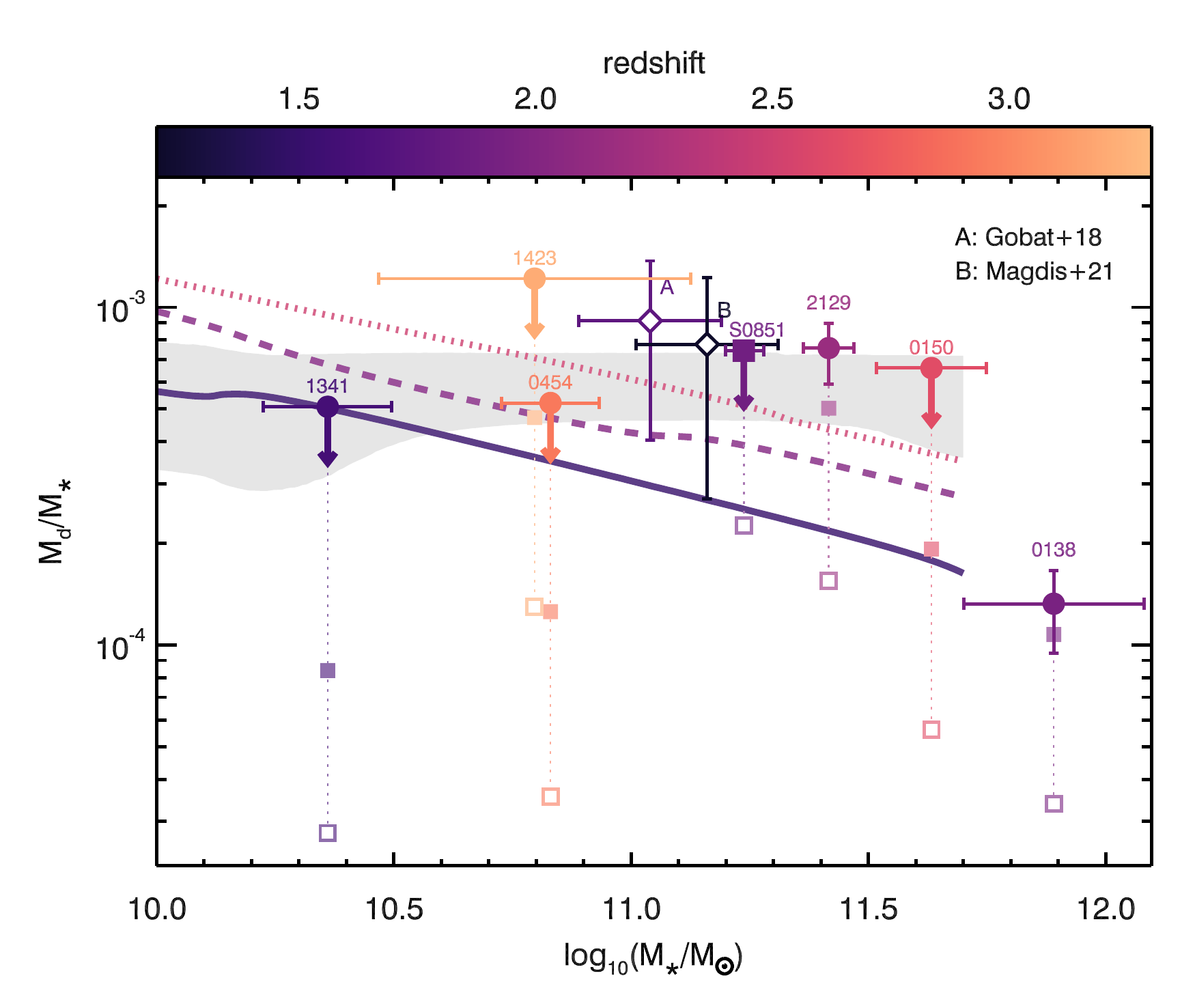}
\caption{Dust-to-stellar mass fraction of lensed QGs \citep[W21;][]{Cal21} as a 
function of redshift (left) and stellar mass (right), compared to stacked samples 
(G18 and M21) and local QGs \citep{Lia16}. 
Filled dots show the constraints derived from extended model fits, while open 
and filled squares correspond to point-source fluxes converted to dust masses 
using the S16 parameterization and M21 templates, respectively. All symbols 
are color-coded as a function of either stellar mass (left panel) or redshift 
(right panel). 
The right panel also shows the expected average dust fraction of QGs, from a 
simple model based on the evolution of their mass function \citep{Gob20}, assuming 
that either the initial dust fraction after quenching is a fixed fraction of the 
main sequence one (solid, dashed, and dotted lines corresponding to $z=1.5$, 2, 
and 2.5, respectively) or constant irrespective of redshift and mass (gray envelope, 
for the same redshift range). In the left panel,  the green curve shows the typical 
dust fraction of main sequence galaxies \citep{Whi14}, while the gray envelope 
corresponds to the redshift evolution of the \citet{Gob20} model.
}
\label{fig:fdust}
\end{figure*}

Under the assumption that interstellar dust and stars have comparable distributions 
in the source plane, the dust-to-stellar mass ratio $M_d/M_{\star}$~does not depend 
on the lensing magnification $\mu$. 
We note that, in any other case, correctly estimating $M_d/M_{\star}$~would 
require the usage of a full lensing model to account for differential 
magnification across images. 
We therefore do not include the uncertainty on $\mu$~in the formal error on this 
ratio. Likewise, here, we focus our discussion on dust fractions, so as to not 
depend on the uncertain dust-to-gas conversion factor (92 for M21, 100 for W21). 
Fig.~\ref{fig:fdust} shows the $M_d/M_{\star}$ of the REQUIEM-ALMA galaxies, as 
a function of redshift and stellar mass, for the three different flux extraction 
and 1.3\,mm-to-dust conversion cases: point sources with the S16 calibration, as 
in W21; point sources with the M21 SED; and extended models with the M21 SED. 
For comparison, we also include in Fig.~\ref{fig:fdust} literature constraints on 
the $M_d/M_{\star}$~of S0851, which is another lensed QG (though unresolved in the 
FIR) discussed in \citet{Cal21}. 
The first case yields the lowest $M_d/M_{\star}$,~while the second one produces 
values higher by a factor $3-3.5$. As noted in M21, the difference between these 
two estimates stems from the higher $T_d$~used by S16, which implies a higher 
dust luminosity per unit mass. Using the extended emission model (third case) 
raises $M_d/M_{\star}$~by another 25\%-50\% for detected objects and a factor 
3-5 in the case of upper limits, depending on the size of the galaxy.\\

Using the stellar light model and M21 SED, the constraints on the $M_d/M_{\star}$ 
of the REQUIEM-ALMA galaxies become consistent with those derived from FIR 
SEDs (G18, M21) and thus with a scenario where relatively young QGs retain 
a non-negligible ISM, as found by \citet{Sue17} and \citet{Bez22} at 
intermediate redshift.
However, we caution that, while at least one object (MRG-2129) has 0.13\% dust, 
implying $f_{\text{gas}}\sim10$\%, the upper limits derived here remain 
compatible with much lower ISM fractions, which makes it difficult to draw firm 
conclusions on quenching scenarios from these data.\\
 
The notable exception is MRG-0138 which, even in this new analysis, 
remains almost an order of magnitude more dust-poor than the other five QGs. 
Assuming no systematic error in its estimated magnification, it is also the 
most massive galaxy in the sample, occupying the high-mass tip of the galaxy 
mass function in its redshift range \citep[e.g.,][]{Dav17}, 
and it is possibly the oldest when accounting for differences in redshift 
(N18 and Appendix~\ref{appendix:muse}). This suggests a host dark matter halo 
with a mass $>$$5\times10^{13}$\,M$_{\odot}$ 
\citep[at 3$\sigma$, assuming parameter and $M_{\star}$ uncertainties as well 
as the $M_{\star}-M_{\text{halo}}$~relation of][]{Gir20}, where cold gas 
accretion should have become inefficient at $z\gtrsim3$ \citep{DB06}, around 
which time quenching likely began. 
Since dust is destroyed over time in QGs \citep[e.g.,][]{Sme18,Whi21b}, its 
low concentration in MRG-0138 is therefore unsurprising. For example, a depletion 
time for dust of $\sim$1.7\,Gyr \citep{Mic19,Gob20} would imply that the initial 
dust mass after quenching was higher by at least a factor 2. This mirrors 
the anticorrelation between $f_{\text{gas}}$~and stellar mass seen in the 
local Universe \citep[e.g.,][]{You11,Sai17,Tac18,Tac20,Liu19,Sai22}.

\section{\label{conclusion}Conclusions}

In this Letter, we revisited the REQUIEM-ALMA literature sample of six lensed QGs. 
We analyzed the ALMA Band 6 continuum observations of these galaxies, described 
in W21, with a different methodology and assumptions. Namely, we allowed for extended 
dust emission coterminous with the stellar distribution, with or without an 
additional compact component, and we converted the measured intensities to dust 
masses using an empirical FIR SED derived from larger statistical QG samples. 
We found that, while some tension with SED-derived results might still exist, 
the method used for FIR flux extraction and the light-to-mass conversion 
factor both have a significant impact on the estimate of their dust fractions, 
with both adding up to an order of magnitude variation. In particular, opting 
for extended models based on the stellar surface brightness of the targets has 
the effect of relaxing, if not eliminating, said tension. 
Under our physically motivated assumptions, the constraints on the dust fraction 
of this sample are broadly consistent with the $M_d/M_{\star}\sim0.08-0.09$\% 
values derived by G18 and M21 for coeval QGs from lower-resolution FIR stacks.\\

On the other hand, we confirm the scant dust content of MRG-0138 determined 
in W21. This object, being the most massive member of the sample, has likely 
experienced the most biased evolution. Its tension with the stacked results 
of G18 and M21 might then be explained by an anticorrelation between ISM 
mass and stellar mass being already present at this epoch, which would be 
largely diluted with them being averaged over large samples with relatively 
broad stellar mass ranges.\\

Knowing whether the ISM of high-redshift QGs is confined to central cores, 
distributed but clumpy, or diffuse would drive the choice of the extraction 
method. Simulations could in principle inform us on the likely true amount 
and distribution of dust and gas in early QGs, if sufficiently well bracketed 
by other observables. On the other hand, the current statistics allowed by 
strong lensing, and the data thereon at their current depth, are not 
sufficient to draw definitive conclusions on this aspect. Given the 
current lack of consensus regarding the total ISM mass of these 
galaxies, resolved observations thus need to be taken with caution.

\begin{acknowledgements}

GBC acknowledges the Max Planck Society for financial support through the Max 
Planck Research Group for S.H.Suyu and the academic support from the German 
Centre for Cosmological Lensing.

\end{acknowledgements}

\begin{appendix}

\section{\label{appendix:photometry}HST and Spitzer coverage}

Table~\ref{tab:hst} shows the broadband photometric coverage by the HST 
and \emph{Spitzer} telescopes of the fields of each QG, with each dot indicating 
that public imaging data are available from their respective archives.   

\begin{table}[h]
\caption{Broadband photometric coverage}
\resizebox{0.49\textwidth}{!}{
\begin{tabular}{c c c c c c c}
\hline\hline
\centering
Band & \multicolumn{6}{c}{MRG-}\\
& 0138 & 0150 & 0454 & 1341 & 1423 & 2129\\
\hline
F275W & & & & & $\bullet$ & \\
F336W & & & & & $\bullet$ & $\bullet$\\
F390W & $\bullet$ & & & & $\bullet$ & $\bullet$\\
F435W & & & & & $\bullet$ & $\bullet$\\
F475W & & & & & $\bullet$ & $\bullet$\\
F555W & $\bullet$ & & $\bullet$ & & $\bullet$ & $\bullet$\\
F606W & & $\bullet$ & & $\bullet$ & $\bullet$ & $\bullet$\\
F625W & & & & & & $\bullet$\\
F775W & & & $\bullet$ & & $\bullet$ & $\bullet$\\
F814W & $\bullet$ & $\bullet$ & $\bullet$ & $\bullet$ & $\bullet$ & $\bullet$\\
F850LP & & & $\bullet$ & & $\bullet$ & $\bullet$\\
F105W & $\bullet$ & $\bullet$ & $\bullet$ & $\bullet$ & $\bullet$ & $\bullet$\\
F110W & $\bullet$ & $\bullet$ & $\bullet$ & $\bullet$ & $\bullet$ & $\bullet$\\
F125W & $\bullet$ & $\bullet$ & $\bullet$ & $\bullet$ & $\bullet$ & $\bullet$\\
F140W & $\bullet$ & $\bullet$ & $\bullet$ & $\bullet$ & $\bullet$ & $\bullet$\\
F160W & $\bullet$ & $\bullet$ & $\bullet$ & $\bullet$ & $\bullet$ & $\bullet$\\
IRAC1 & $\bullet$ & $\bullet$ & $\bullet$ & $\bullet$ & $\bullet$ & $\bullet$\\
IRAC2 & \phantom{$\bullet$} & $\bullet$ & $\bullet$ & $\bullet$ & $\bullet$ & $\bullet$\\
IRAC3 & & & $\bullet$ & & $\bullet$ & $\bullet$\\
IRAC4 & & & $\bullet$ & & $\bullet$ & $\bullet$\\
\hline
\end{tabular}
}
\label{tab:hst}
\end{table}

\section{\label{appendix:muse}Spatially resolved spectroscopy}

We obtained integral field spectroscopic data of two of the 
six targets from the Science Archive of the European Southern 
Observatory (ESO). The clusters MACS0138 and MACS1341 were observed
with the Multi Unit Spectroscopic Explorer (MUSE) on the Very Large 
Telescope (VLT) under ESO programs 0103.A-0777 (PI A. Edge) and 
0103.B-0717 (PI A. Man), respectively. 
MACS0138 was observed for a total of 48.5 minutes on target with a 
median seeing of $\sim$0.8'', while MACS1341 was observed for a total
of 82 minutes using the Ground Layer Adaptive Optics (GLAO) system, 
which provided a final point spread function (PSF) width of 0.55'' 
measured from the white light image (i.e., the data cube collapsed 
along its wavelength axis). The raw data were reduced using the 
standard MUSE pipeline \citep[version 2.8;][]{MUSE} to produce 
corrected and calibrated data cubes. We then applied the {\tt Zurich 
Atmosphere Purge} tool \citep[ZAP;][]{ZAP} to improve the sky 
subtraction and, for each cluster, matched the world coordinate 
system (WCS) of the data cube (based on the white image) to that 
of the HST/WFC3 F160W image. Further details on the data and their 
reduction will be given in Caminha et al. (in prep.).\\

We divided each of the F160W stellar light models into two areas of 
equal total flux, corresponding to an inner (or ``bulge'') and outer 
(or ``disk'') region. We then rebinned these masks to the MUSE 
pixel size and produced a median stacked 1D spectrum for each. 
We computed an associated noise spectrum using the jackknife method 
as 
\begin{equation}\label{eq:jackknife}
\sigma = \sqrt{\frac{N-1}{N}\sum_{i=1}^{N}(f-f_i)^2}~\text{,}
\end{equation}
where $N$~is the number of pixels, $f_i$~the spectrum corresponding 
to the $i$-th pixel, and $f$~the coadded spectrum.\\

\begin{figure}[h]
\centering
\includegraphics[width=0.499\textwidth]{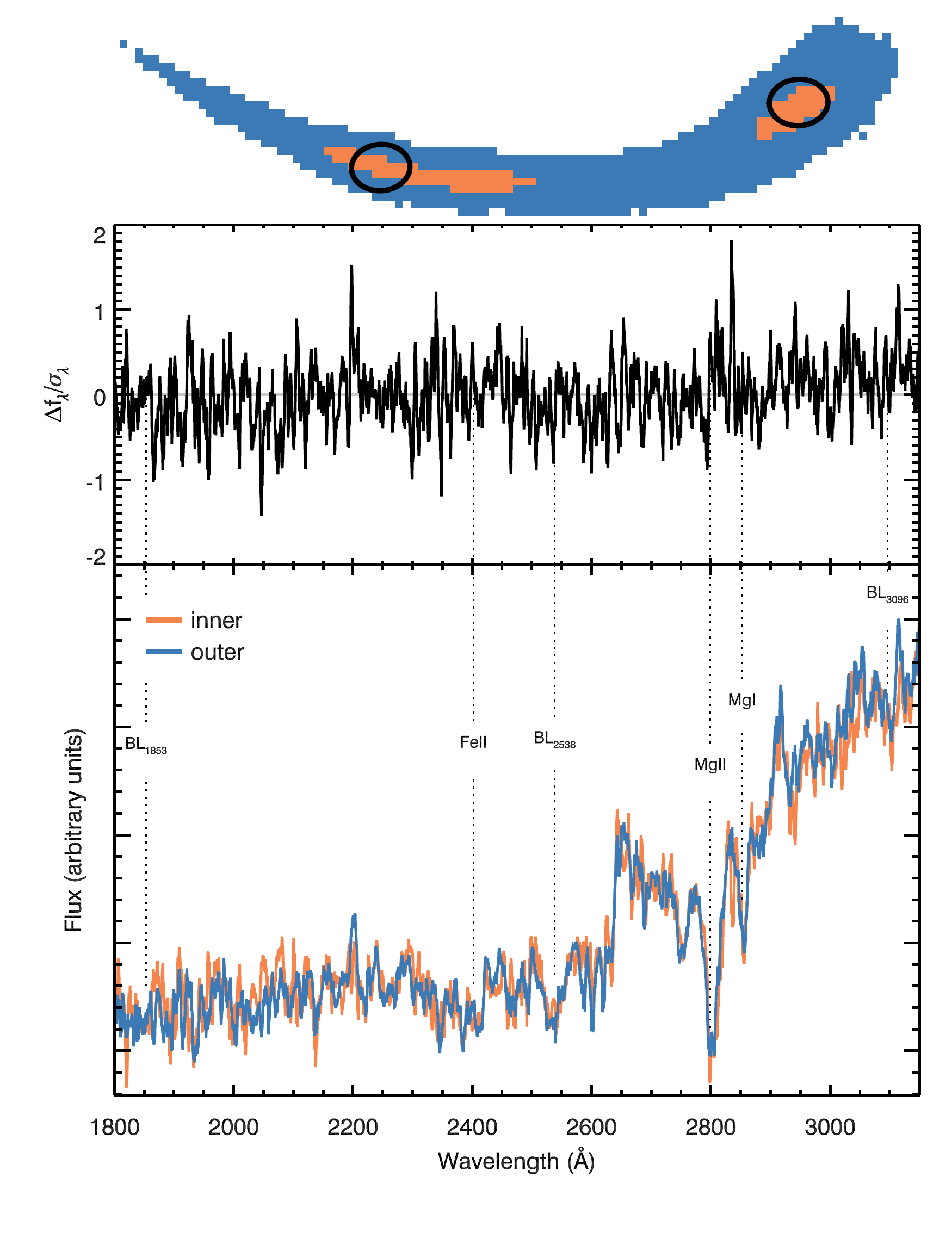}
\caption{\emph{Bottom panel}: Average VLT/MUSE spectra of inner (orange) 
and outer (blue) regions in the MRG-0138 arc image, covering the 
$1800-3150$\,\AA~rest-frame wavelength range and smoothed with a ten-pixel 
kernel for clarity. Significant near-ultraviolet features are indicated 
by dotted lines.
\emph{Middle panel}: Signal-to-noise ratio of the difference spectrum 
(inner $-$ outer) as a function of wavelength.
\emph{Top panel}: Mask of the MRG-0138 arc showing inner (orange) and outer 
(blue) regions, with the corresponding ALMA band 6 beams shown as black 
ellipses.
}
\label{fig:m0138}
\end{figure}

\begin{figure}[h]
\centering
\includegraphics[width=0.499\textwidth]{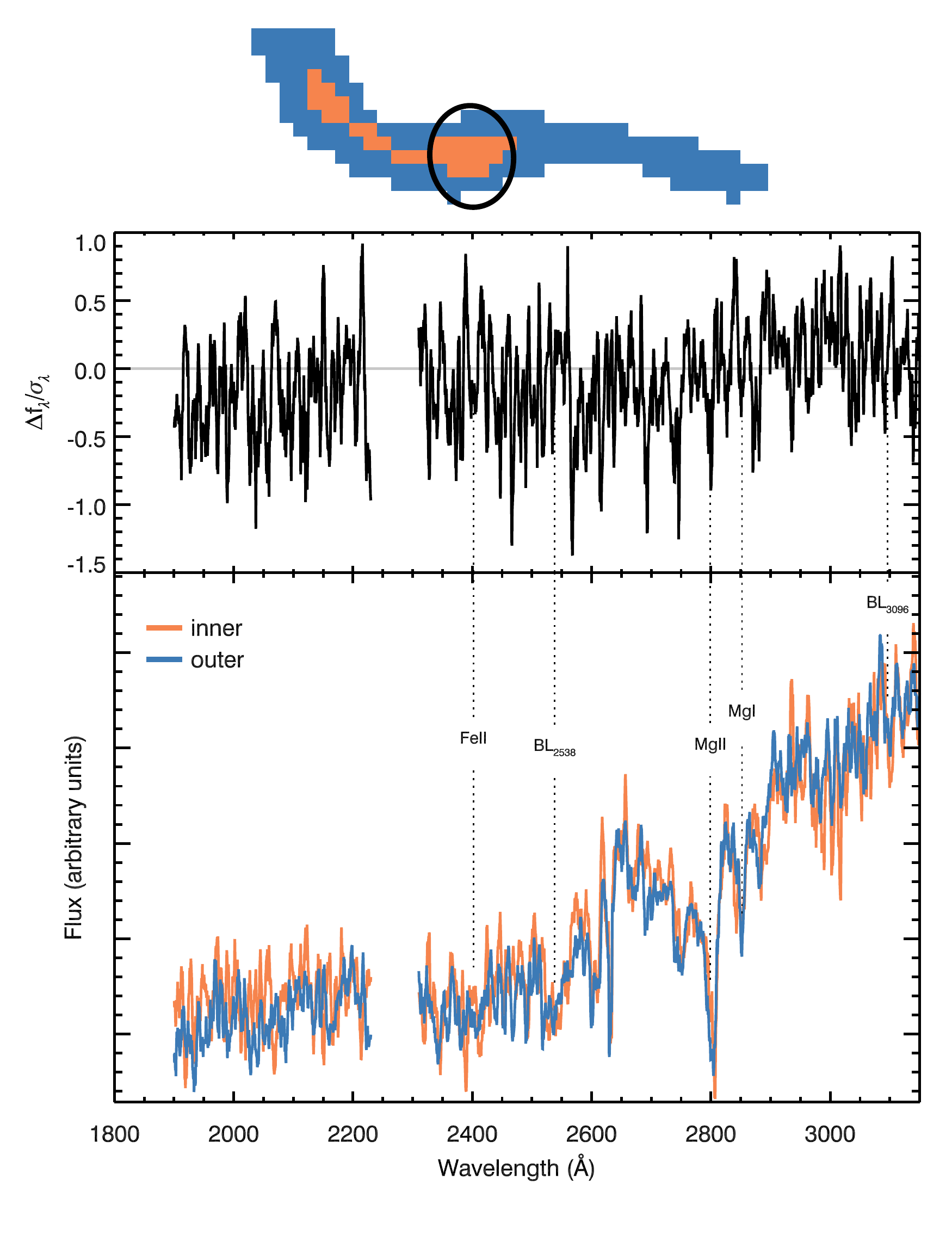}
\caption{Same as Fig.~\ref{fig:m0138}, but for MRG-1341.
}
\label{fig:m1341}
\end{figure}

Figs.~\ref{fig:m0138} and \ref{fig:m1341} compare the inner and outer 
VLT/MUSE spectra of MRG-0138 and MRG-1341, respectively, with 
the signal-difference-to-noise spectrum shown in the middle panel. 
In both cases, we find no appreciable difference between spectral 
features within the two regions. Spectral modeling based on the 
parametric SFH described in Appendix~\ref{appendix:masses} produces 
consistent mass-weighted ages of $\sim$1.5\,Gyr and $\sim$1.8\,Gyr. Similarly, adopting a nonparametric fit with \citet{MILES} 
templates, as done in \citet{Gob17}, yield very similar age and 
metallicity distributions, within uncertainties. 
However, we caution that, although the absorption features covered 
by the MUSE spectra correlate with a combination of age and metallicity 
\citep{Fan92,Mar09}, being limited to the near-ultraviolet, which 
is most sensitive to young massive stars, makes them suboptimal for 
reconstituting SFHs. In addition, some stellar populations could be 
invisible if, for example, they are entirely shrouded by dust. Alternatively, 
a signal of inside-out or outside-in quenching might still be present, 
but more subtle than what these particular data can resolve. 
Finally, we note that these results do not change if mean spectra 
are used or if we define the inner region using the ALMA beam (see top 
panels in Figs.~\ref{fig:m0138} and \ref{fig:m1341}) rather than 
isophotes.

\section{\label{appendix:fit}\emph{uv} plane modeling}

Given an intensity distribution on the sky of $M(x,y)$, where $x$~and $y$~are 
spatial coordinates from the phase center, we modeled the complex visibility 
as\\
\begin{equation}\label{eq:vis}
V(u,v) = \iint B(x,y)M(x,y)e^{2\pi i(ux+vx)}\mathrm{d}x\mathrm{d}y~\text{,}
\end{equation}
\noindent
where $B(x,y) = (2J_1(\theta)/\theta)^2$~is the primary beam response of the 
antennas, with $J_1$ the first order Bessel J function,  
$\theta=\sqrt{x^2+y^2}/\text{HWBN}$, and HWBN is the half-beam width at 
first null. 
Here we assumed that there are no significant differences in astrometry 
between the ALMA and HST data. In cases where a bright off-center 
source exists in the field of view (e.g., for MRG-1341; see 
Fig.~\ref{fig:cutouts}), this appears to be the case.

\section{\label{appendix:masses}Comparison of stellar masses}

\begin{figure}[h]
\centering
\includegraphics[width=0.499\textwidth]{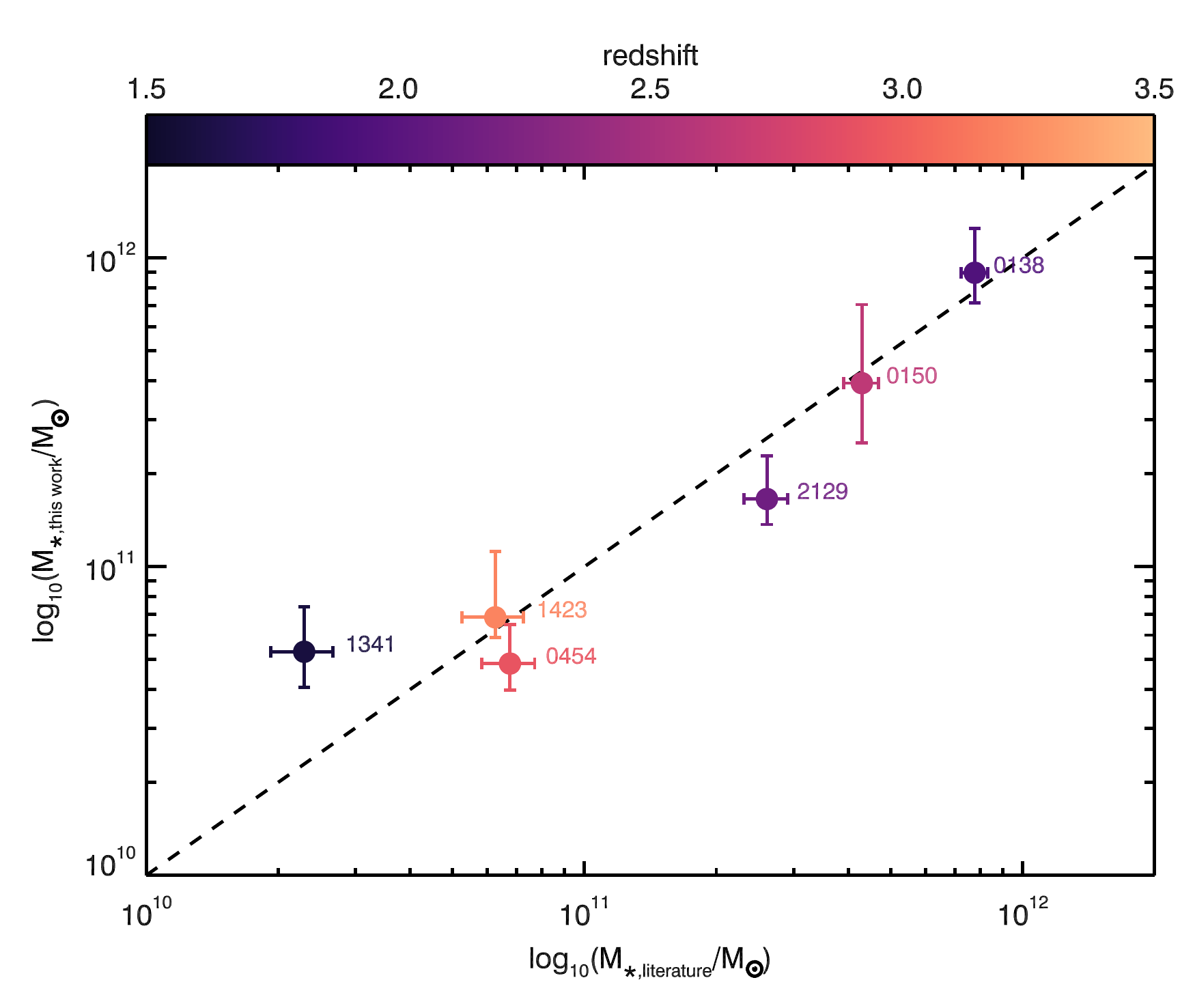}
\caption{Comparison between stellar masses of the REQUIEM-ALMA sample 
reported in previous works \citep{New18,Man21} and computed here from 
HST+\emph{Spitzer} SEDs with metallicity priors. The masses 
from the literature were converted to a \citet{Sal55} IMF and both 
have been de-magnified; however, the error bars shown here do not 
account for magnification uncertainties. The dashed line shows the 
one-to-one relation.
}
\label{fig:masses}
\end{figure}

Fig.~\ref{fig:masses} compares previously published stellar masses of the 
QG sample with those used in this work. The latter masses were estimated 
from fits to their HST and \emph{Spitzer} SEDs, extracted using SExtractor 
\citep{BA96} in dual-image mode with the HST/WFC3 F160W model as the base, 
using \citet{BC03} stellar population models. We assumed metallicity priors 
from spectroscopy \citep{New18,Man21} and delayed exponentially declining 
SFHs with an exponential cutoff of the form
\begin{equation}\label{eq:SFH}
\text{SFR}(t)\propto\frac{t}{\tau^2}e^{-t/\tau}\text{min}(e^{-(t-t_q)/\tau_q},1)
\end{equation}
(where $t$~is the time since the beginning of star formation, $\tau$~the SFH's 
$e$-folding timescale, and $t_q$~and $\tau_q$~parameterize the 
start and timescale of quenching, respectively), which is appropriate for 
high-redshift QGs \citep[e.g.,][]{Sch18,Val20}. 
Here we fixed the start of the SFH at $z=10$~and assumed parameter ranges of 
$(0,3]$~Gyr for $\tau$, $[1,3]$~Gyr for $t_q$, and $[0.1,1.]$~Gyr for $\tau_q$.\\

The stellar masses thusly derived from the SEDs are consistent with those 
published in previous works and used in W21, except for MRG-1341, where our 
estimate is almost a factor 2 higher. As a consequence, the $M_d/M_{\star}$ 
of MRG-1341 presented in this Letter is lower by half than if we had used 
the literature value.

\end{appendix}

\end{document}